\documentclass[preprint,5p,times,twocolumn]{elsarticle}
\usepackage{lineno,hyperref}
\usepackage{hyperref}
\usepackage{subcaption}
\usepackage{xcolor}
\usepackage{booktabs}
\usepackage{multirow}
\usepackage{float}

\usepackage{graphicx,color,amsmath,amssymb,chemformula}

\begin{document}
\begin{frontmatter}
\title{Spin Response Properties in Electronically Robust Ferromagnetic Strained $\text{CrSiSe}_3$ Monolayer under External Electric Fields}

\author{S Solihin$^{\text{a,b,}*}$}
\ead{solihin@ui.ac.id}

\author{Ahmad R. T. Nugraha$^{\text{b,c}}$}
\ead{ahma080@brin.go.id}

\author{Muhammad Aziz Majidi$^{\text{a}}$}
\ead{aziz.majidi@sci.ui.ac.id}

\address{
\vspace{1mm}
$^{a}$Department of Physics, Faculty of Mathematics and Natural Sciences, Universitas Indonesia, Depok 16424, Indonesia\\
$^{b}$Research Center for Quantum Physics, National Research and Innovation Agency (BRIN), South Tangerang 15314, Indonesia\\
$^{c}$Engineering Physics Study Program, School of Electrical Engineering, Telkom University, Bandung 40257, Indonesia
}

\begin{abstract}
Integrating two-dimensional van der Waals magnets into field-effect spintronic devices requires robust charge stability and tunable spin responses.  In this study, we investigate the electronic, topological, magnonic, and magneto-optical properties of the strain-engineered ferromagnetic $\text{CrSiSe}_3$ monolayer under out-of-plane external electric fields by using first-principles calculations.  We find that for this material, the intrinsic charge properties, including the indirect band gap, charge Berry curvature, optical conductivities, and magneto-optical Kerr effect spectra, exhibits exceptional robustness against applied fields up to 0.3~V/\AA.  Conversely, the spin degrees of freedom demonstrate highly sensitive tunability.  Electrostatic gating significantly modulates the spin Berry-like curvature, driving a non-monotonic enhancement in the spin Hall conductivity.  Furthermore, external fields effectively tune collective magnon excitations by modifying microscopic Heisenberg exchange interactions.  Such coexistence of charge immunity and flexible spin manipulation establishes the strained $\text{CrSiSe}_3$ monolayer as a promising platform for stable spintronic devices.
\end{abstract}

\end{frontmatter}

\section{Introduction}
\label{sec:introduction}

The discovery of intrinsic two-dimensional (2D) magnetism in van der Waals (vdW) materials has sparked a kind of ``revolution'' in condensed matter physics and materials science \cite{park2026vdw,hwang2026van,novoselov20162d} since, unlike their bulk counterparts, 2D vdW magnets offer unprecedented opportunities to scale down information storage and processing devices to the atomic limit \cite{chen2024van,ding2025recent}.  In particular, monolayer magnetic thin films serve as ideal and highly promising platforms for advanced magnonic and spintronic applications \cite{sundararajan2025toward,chen2021electrically,sierra2021van,roche2024spintronics,ahn20202d}. In these nanoscale architectures, spin waves (magnons) and spin-polarized currents are utilized to transmit and process information, offering a paradigm shift towards ultra-low power consumption and minimal Joule heating compared to conventional charge-based electronics \cite{mook2015magnon,wang2024nanoscale,wang2023magnon,chumak2015magnon}.

Among the diverse families of 2D vdW magnets, the chromium-based ternary chalcogenides, denoted as Cr$M$$X_3$ (where $M =$ Si, Ge and $X =$ S, Se, Te), have attracted significant attention due to their intrinsic semiconducting nature and rich magnetic phase diagrams. Extensive theoretical and experimental efforts have been dedicated to exploring the prototypical members of this family, such as $\text{CrGeTe}_3$ and $\text{CrSiTe}_3$, revealing fascinating magneto-optic and topological phenomena \cite{zhu2021topological,chen2022anisotropic,wang2020magnon}.  However, despite sharing a similar structural framework, the selenium-based counterpart, $\text{CrSiSe}_3$, remains notably under-explored in the current literature \cite{shen2020first,yu2025magnetic}.  This distinct lack of comprehensive studies presents a knowledge gap, particularly regarding its dynamic spin responses, magnonic excitations, and topological properties under external perturbations.  Bridging this gap is essential, especially when considering the stringent operational requirements for practical device integration. 

To actively manipulate the magnetic states in 2D materials, applying an external electric field via electrostatic gating is highly preferred over conventional magnetic fields or chemical doping. Electric field gating offers non-volatile, highly localized control, and seamless compatibility with modern complementary metal-oxide-semiconductor (CMOS) architectures. However, subjecting 2D materials to such external voltages introduces a critical challenge. The material's fundamental electronic properties must exhibit a high degree of stability to prevent unwanted dielectric breakdown or unintended metallic transitions when stray electric fields or required gate voltages are applied in realistic environments.  Consequently, for successful implementation in modern field-effect spintronic and magnonic devices, a candidate material must satisfy a complex, dual requirement.  While the fundamental electronic band gap and charge transport mechanisms must be strictly protected to ensure operational stability, the spin degrees of freedom must simultaneously remain highly responsive. This selective tunability is critical to allow for the active modulation of spin currents, spin Hall conductivities, and magnon dispersions via electrical gating \cite{tasiu2025spintronics,liu2025strain,dieny2020opportunities}. 
Identifying 2D materials that maintain robust charge properties while enabling tunable spin responses is quite an active research direction.

Motivated by application-driven requirements mentioned above, in this work, we investigate the spin response properties of the $\text{CrSiSe}_3$ monolayer under out-of-plane external electric fields using first-principles calculations.  While previous computational studies on this material primarily relied on collinear spin approximations, it should be noted that investigating dynamic spin responses, topological invariants, and macroscopic spin Hall conductivity fundamentally requires the rigorous inclusion of spin-orbit coupling (SOC) within a noncollinear framework.  
Furthermore, since the unstrained $\text{CrSiSe}_3$ naturally exhibits an antiferromagnetic ground state, we apply uniform biaxial tensile strains of 2.5\% and 5\% to securely stabilize the ferromagnetic phase, establishing the necessary magnetic baseline for active spintronic applications.  By comprehensively evaluating the electronic band structures, charge and spin topological invariants, frequency-dependent optical conductivities, and magnon dispersions across these strain configurations, we uncover an important dichotomy in the strained $\text{CrSiSe}_3$ monolayer.  In particular, its fundamental charge properties and optical transport signatures are immune to out-of-plane electric fields, whereas its spin Hall conductivity and macroscopic magnon excitations demonstrate highly responsive and controllable tunability.  We systematically elaborate our model, methods, and results below, followed by conclusions and further remarks.

\section{Material Model and Computational Methods}
\label{sec:model}

\subsection{\texorpdfstring{$\text{CrSiSe}_3$}{CrSiSe3} Monolayer Structure}
\label{subsec:structure}

\begin{figure}[t]
    \centering
    \includegraphics[width=\linewidth]{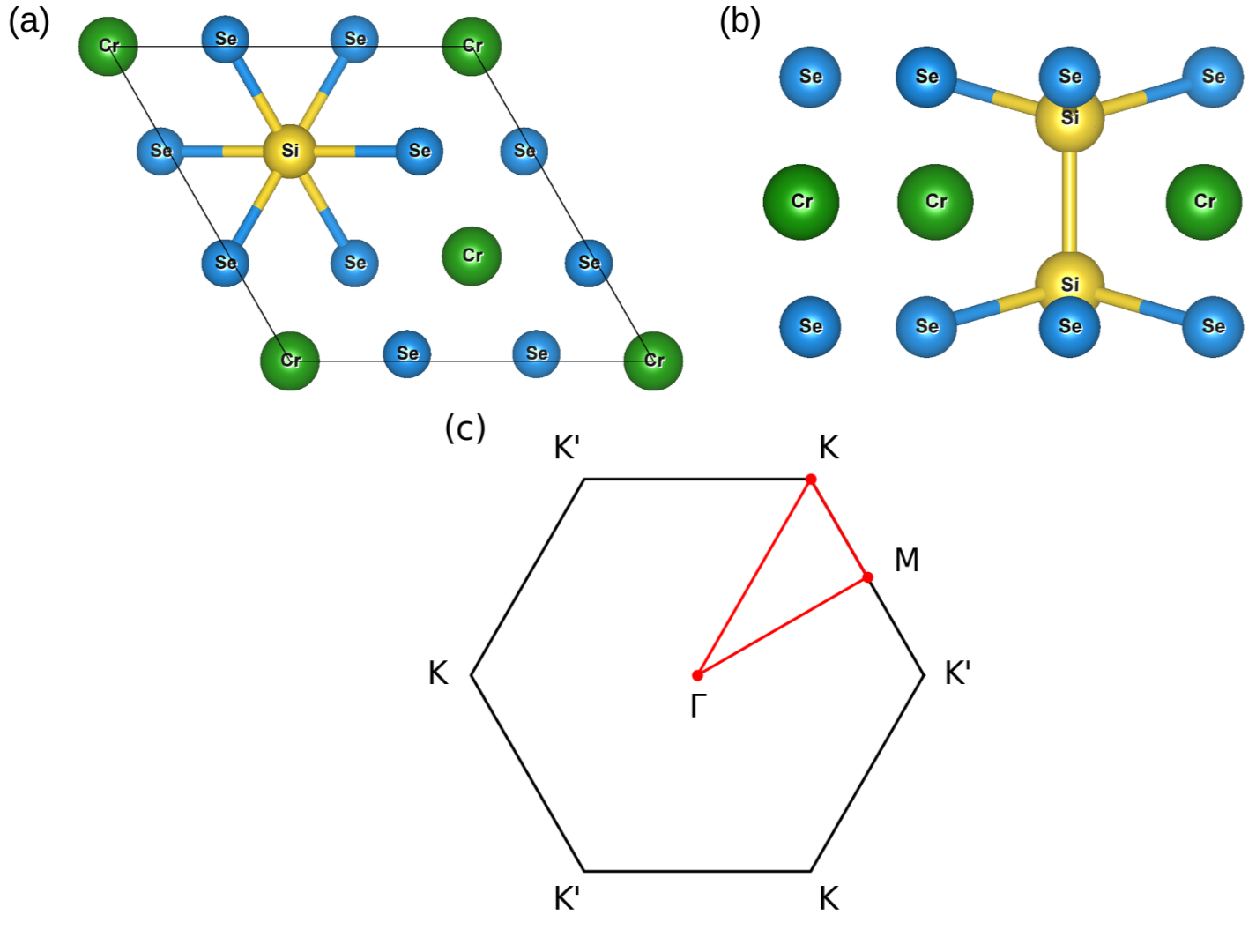}
    \caption{Optimized geometrical structure of the unstrained $\text{CrSiSe}_3$ monolayer. Panels (a) and (b) display the top and side views of the atomic configuration, respectively. Panel (c) illustrates the corresponding 2D Brillouin zone along with the selected high-symmetry $\mathbf{k}$-path ($\Gamma$-M-K-$\Gamma$) utilized for the momentum-resolved calculations.}
\label{fig:structure}
\end{figure}

The atomic configuration of the $\text{CrSiSe}_3$ monolayer is modeled within a hexagonal unit cell. To ensure the physical isolation of the 2D layer and strictly eliminate any spurious interactions between adjacent periodic images, the total length of the cell along the $z$-axis is set to 25~\AA\ with the monolayer centrally positioned within the vacuum space. The optimized geometric structure is illustrated in Fig.~\ref{fig:structure}. Panels (a) and (b) of Fig.~\ref{fig:structure} display the top and side views of the atomic arrangement, while panel (c) depicts the corresponding 2D Brillouin zone alongside the high-symmetry $\mathbf{k}$-path ($\Gamma$-M-K-$\Gamma$).

In its unstrained phase, the optimized in-plane lattice constant of the $\text{CrSiSe}_3$ monolayer is found to be $a_0 = 6.28$ \AA \cite{shen2020first}. To investigate the effects of mechanical deformation, a uniform biaxial tensile strain ($\varepsilon$) is applied to the structural model. The strained lattice parameter $a$ is quantitatively defined by the relation $a = a_0(1 + \varepsilon)$. By imposing representative strain magnitudes of $\varepsilon = 2.5\%$ and $\varepsilon = 5\%$, the in-plane lattice constant is isotropically expanded to $a \approx 6.44$ \AA\ and $a \approx 6.60$ \AA, respectively. Following these lattice expansions, the internal atomic coordinates are completely relaxed, while the strained cell dimensions are kept fixed to preserve the applied strain conditions.

The magnetic behavior of $\text{CrSiSe}_3$ is primarily governed by the competition between various exchange interaction ($J$) mechanisms. Specifically, the magnetic ground state emerges from the direct exchange and superexchange interactions between the nearest-neighbor (NN) Cr atoms, supplemented by the superexchange interactions involving the second and third NN sites. According to previous studies \cite{shen2020first}, unstrained $\text{CrSiSe}_3$ exhibits an antiferromagnetic (AFM) ground state. However, the material undergoes a magnetic phase transition to a ferromagnetic (FM) state when subjected to a tensile strain exceeding 2\% \cite{yu2025magnetic,shen2020first}. Based on such reported phase transitions, our work here focuses on the 2.5\% and 5\% strained structures, which effectively stabilize the FM state.

\subsection{Computational Methods}
\label{comp}

We perform first-principles calculations based on density functional theory (DFT) using the Quantum ESPRESSO package~\cite{giannozzi2009quantum}. The exchange-correlation potential is treated with the Perdew-Burke-Ernzerhof (PBE) functional within the generalized gradient approximation (GGA)~\cite{perdew1996gga}. To describe the ion-electron interactions, we employ the optimized norm-conserving Vanderbilt (ONCV) pseudopotentials \cite{hamann2013optimized} obtained from PseudoDojo~\cite{van2018pseudodojo}.  Unlike previous computational studies on this material that were restricted to collinear approximations \cite{shen2020first,yu2025magnetic}, the spin-orbit coupling (SOC) is explicitly incorporated through a noncollinear formalism.  This rigorous treatment is essential to accurately capture the topological invariants, spin transport signatures, and anisotropic exchange interactions. 

To account for long-range dispersion forces, van der Waals (vdW) interactions are included using the DFT-D3 method \cite{grimme2010consistent}. Furthermore, to address the strong on-site Coulomb interactions among the localized Cr-$3d$ electrons, we apply the DFT+$U$ \cite{dudarev1998electron} scheme with an effective $U$ value of 4.0 eV \cite{li2020large}. The electronic wavefunctions are expanded in a plane-wave basis set with a kinetic energy cutoff of 100 Ry and Brillouin zone integration is performed using a $12 \times 12 \times 1$ Monkhorst-Pack $k$-point grid~\cite{monkhorst1976special}. To simulate the effect of an external electric field, a sawtooth potential is applied along the direction perpendicular to the monolayer plane~\cite{hung2022quantum}.

The converged Bloch eigenstates are subsequently used to construct maximally localized Wannier functions (MLWFs) via the Wannier90 package \cite{pizzi2020wannier90}, utilizing the projections of the Cr-$d$, Si-$p$, and Se-$p$ orbitals. To obtain the topological and optical signatures in reasonable accuracy, the Wannier-interpolated tight-binding Hamiltonian is evaluated over an ultra-dense $k$-point mesh of $250 \times 250 \times 1$.   We compute the intrinsic Berry curvature in particular to evaluate the topological characteristics.  In terms of the cell-periodic Bloch states $\vert u_{n\mathbf{k}}\rangle$, the Berry curvature for the $n$-th energy band is defined as the curl of the Berry connection \cite{xiao2010berry,wang2006ab},
\begin{equation}
\label{eq:berry_curv}
\boldsymbol{\Omega}_n(\mathbf{k}) = \boldsymbol{\nabla}_{\mathbf{k}} \times \langle u_{n\mathbf{k}}\vert i\boldsymbol{\nabla}_{\mathbf{k}} \vert u_{n\mathbf{k}}\rangle.
\end{equation}
For a rigorous evaluation within our computational scheme, the Berry curvature is expressed as an antisymmetric tensor,
\begin{equation}
\label{eq:curv_tensor}
\Omega_{n,\alpha\beta}(\mathbf{k}) = \epsilon_{\alpha\beta\gamma} \Omega_{n,\gamma}(\mathbf{k}) = -2\text{Im}\langle \nabla_{k_\alpha} u_{n\mathbf{k}}\vert \nabla_{k_\beta} u_{n\mathbf{k}}\rangle.
\end{equation}
For a 2D system like the $\text{CrSiSe}_3$ monolayer, the anomalous transport properties are dictated by the out-of-plane component of the Berry curvature, $\Omega_z(\mathbf{k})$. By employing the Wannier interpolation scheme, the total charge Berry curvature is evaluated by summing the contributions over all occupied valence bands. 

Furthermore, we calculate the frequency-dependent optical conductivity tensor, $\sigma_{\alpha\beta}(\omega)$, to evaluate the macroscopic optical responses.  This conductivity tensor is computed by utilizing the Kubo-Greenwood formalism within the independent-particle approximation. The general expression for the interband optical conductivity tensor is given by~\cite{yates2007spectral, pizzi2020wannier90}
\begin{equation}
\label{eq:kubo_general}
\sigma_{\alpha\beta}(\omega) = \frac{ie^2\hbar}{V N_k} \sum_{\mathbf{k},n \neq m} \frac{(f_{m\mathbf{k}} - f_{n\mathbf{k}})v_{nm}^\alpha(\mathbf{k}) v_{mn}^\beta(\mathbf{k})}{(\varepsilon_{n\mathbf{k}} - \varepsilon_{m\mathbf{k}})[\varepsilon_{n\mathbf{k}} - \varepsilon_{m\mathbf{k}} - (\hbar\omega + i\eta)]},
\end{equation}
where $V$ is the unit cell volume, $N_k$ is the total number of sampled $\mathbf{k}$-points, and $f_{n\mathbf{k}}$ is the Fermi-Dirac distribution function for the $n$-th band with an energy eigenvalue of $\varepsilon_{n\mathbf{k}}$. The term $v_{nm}^\alpha(\mathbf{k})$ represents the off-diagonal velocity matrix elements along the Cartesian direction $\alpha$, and $\eta$ is a small phenomenological smearing parameter \cite{pizzi2020wannier90}.

To better understand the microscopic mechanisms under light excitation, the total optical conductivity tensor can be decomposed into its Hermitian (dissipative) and anti-Hermitian (reactive) parts, expressed as \cite{yates2007spectral, pizzi2020wannier90}
\begin{equation}
\sigma_{\alpha\beta}(\omega) = \sigma_{\alpha\beta}^{\text{H}}(\omega) + \sigma_{\alpha\beta}^{\text{A}}(\omega).
\end{equation}
By utilizing the Dirac identity to represent the finite lifespan of the states, the Hermitian part $\sigma_{\alpha\beta}^{\text{H}}(\omega)$, which describes the dissipative processes via direct optical absorption, reads
\begin{equation}
\label{eq:hermitian}
\sigma_{\alpha\beta}^{\text{H}}(\omega) = \frac{\pi e^2\hbar}{V N_k} \sum_{\mathbf{k},n \neq m} (f_{n\mathbf{k}} - f_{m\mathbf{k}}) \frac{\text{Re}[v_{nm}^\alpha(\mathbf{k}) v_{mn}^\beta(\mathbf{k})]}{\varepsilon_{m\mathbf{k}} - \varepsilon_{n\mathbf{k}}} \delta(\varepsilon_{m\mathbf{k}} - \varepsilon_{n\mathbf{k}} - \hbar\omega),
\end{equation}
where $\delta$ denotes a broadened delta-function implemented using a broadening profile to achieve high numerical accuracy. Conversely, the anti-Hermitian part $\sigma_{\alpha\beta}^{\text{A}}(\omega)$, which dictates the reactive response of the system under optical frequencies, is given by
\begin{equation}
\label{eq:anti_hermitian}
\sigma_{\alpha\beta}^{\text{A}}(\omega) = \frac{e^2\hbar}{V N_k} \sum_{\mathbf{k},n \neq m} (f_{n\mathbf{k}} - f_{m\mathbf{k}}) \frac{\text{Im}[v_{nm}^\alpha(\mathbf{k}) v_{mn}^\beta(\mathbf{k})]}{(\varepsilon_{m\mathbf{k}} - \varepsilon_{n\mathbf{k}})^2 - (\hbar\omega + i\eta)^2}.
\end{equation}

Within the Wannier interpolation framework, after performing the full summation and averaging over the Brillouin zone, these Hermitian ($\sigma^{\text{H}}$) and anti-Hermitian ($\sigma^{\text{AH}}$) components are assembled into the macroscopic symmetric ($\sigma^{\text{S}}$) and antisymmetric ($\sigma^{\text{A}}$) optical conductivity tensors through the exact relations \cite{yates2007spectral, pizzi2020wannier90}
\begin{equation}
\label{eq:sym_antisym}
\begin{aligned}
\sigma^{\text{S}}_{\alpha\beta} &= \text{Re}[\sigma^{\text{H}}_{\alpha\beta}] + i\text{Im}[\sigma^{\text{AH}}_{\alpha\beta}], \\
\sigma^{\text{A}}_{\alpha\beta} &= \text{Re}[\sigma^{\text{AH}}_{\alpha\beta}] + i\text{Im}[\sigma^{\text{H}}_{\alpha\beta}].
\end{aligned}
\end{equation}

Based on the calculated antisymmetric (Hall) component of the optical conductivity, $\sigma_{xy}(\omega)$, the magneto-optic Kerr effect (MOKE) spectra can be evaluated. For a 2D magnetic material residing on a dielectric substrate, the complex Kerr angle under normal light incidence is determined using the relation established in Refs. \cite{pratama2020circular,adhidewata2023trigonal}:
\begin{align}
    \tan 2\theta_K &\approx 2 \text{Re}\left[ \frac{2 Z_0 \sigma_{xy}^{\text{2D}}(\hbar\omega)}{1 - n^2} \right], \\
    \sin 2\eta_K &\approx 2 \text{Im}\left[ \frac{2 Z_0 \sigma_{xy}^{\text{2D}}(\hbar\omega)}{1 - n^2} \right],
\end{align}
where $Z_0$ is the vacuum impedance, $\sigma_{xy}^{\text{2D}}(\hbar\omega)$ is the 2D sheet Hall conductivity, and $n$ is the refractive index of the underlying substrate. A representative value of $n=2$ is adopted for standard insulating substrates.

\begin{figure*}[t]
    \centering
    \includegraphics[width=\linewidth]{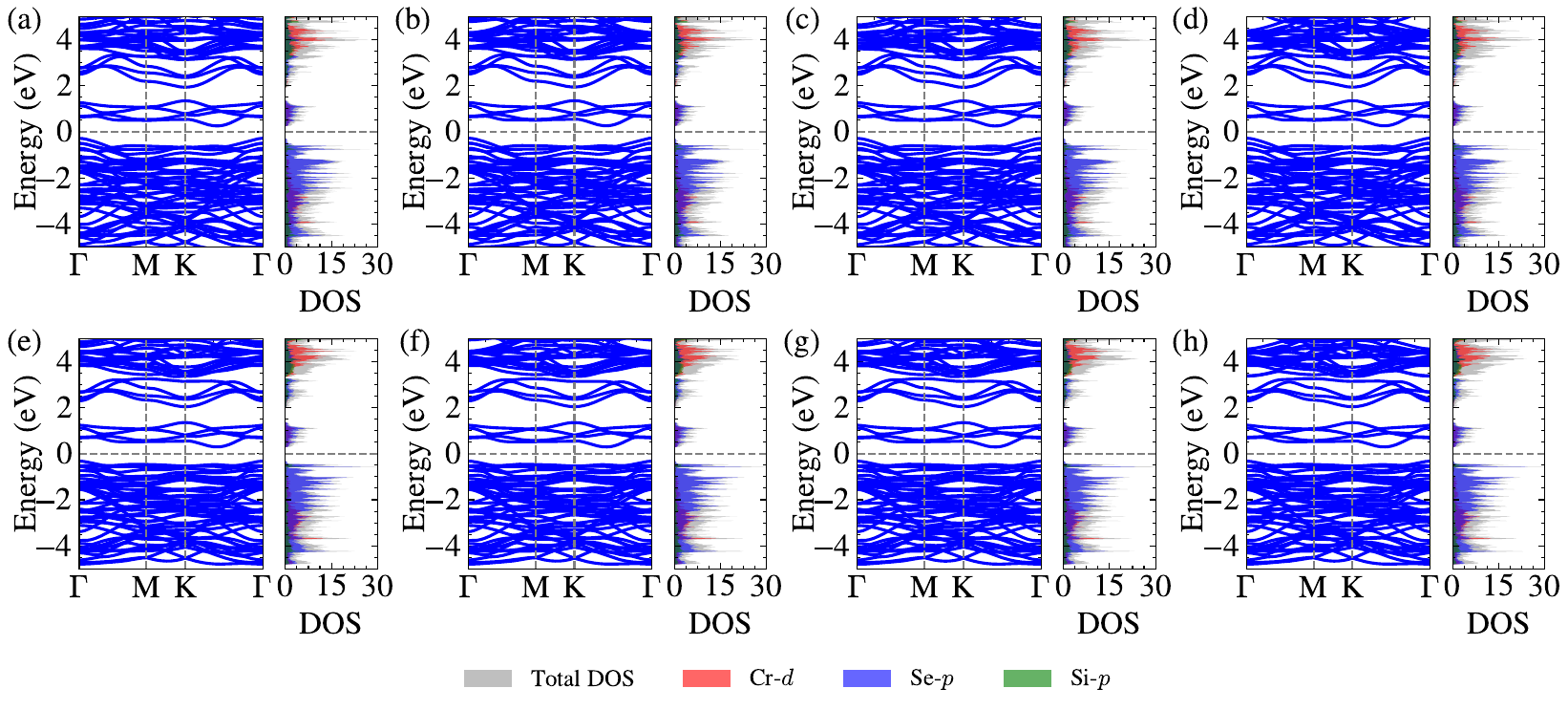}
    \caption{Spin-polarized band structures and the corresponding total density of states (DOS) of the 2.5\% (top row) and 5\% (bottom row) strained $\text{CrSiSe}_3$ monolayer under various out-of-plane external electric fields: (a,e) 0 V/\AA, (b,f) 0.1 V/\AA, (c,g) 0.2 V/\AA, and (d,h) 0.3 V/\AA. The Fermi level is set to zero.}
    \label{fig:band}
\end{figure*}

To capture the spin dynamics, the intrinsic spin Hall conductivity (SHC) is evaluated within the independent-particle approximation using the Kubo-Greenwood formalism \cite{greenwood1958boltzmann}. The spin current operator is defined as $\hat{j}_{\alpha}^{\gamma} = \frac{1}{2}\{\hat{s}_{\gamma}, \hat{v}_{\alpha}\}$, where $\hat{s}_{\gamma} = \frac{\hbar}{2}\hat{\sigma}_{\gamma}$ represents the spin operator and $\hat{v}_{\alpha}$ is the velocity operator. The band-projected spin Berry-like curvature, which acts as the microscopic origin of the SHC, is expressed as
\begin{equation}
\label{eq:spin_berry_band}
\Omega_{n,\alpha\beta}^{\text{spin}\gamma}(\mathbf{k}) = \hbar^2 \sum_{m \neq n} \frac{-2\text{Im}\big[\langle n\mathbf{k}| \frac{1}{2}\{\hat{\sigma}_{\gamma},\hat{v}_{\alpha}\} |m\mathbf{k}\rangle \langle m\mathbf{k}| \hat{v}_{\beta} |n\mathbf{k}\rangle \big]}{(\varepsilon_{n\mathbf{k}} - \varepsilon_{m\mathbf{k}})^2 - (\hbar\omega + i\eta)^2},
\end{equation}
where $\alpha$ and $\beta$ denote the Cartesian directions of the current and electric field, respectively, and $\gamma$ denotes the spin polarization direction (in the studied geometry, $\alpha = x$, $\beta = y$, and $\gamma = z$) \cite{qiao2018calculation,ryoo2019computation,guo2008intrinsic}. The macroscopic spin transport properties are determined by summing these contributions over all occupied bands. The $\mathbf{k}$-resolved spin Berry-like curvature is given by \cite{pizzi2020wannier90,qiao2018calculation}
\begin{equation}
\label{eq:spin_berry_k}
\Omega_{\alpha\beta}^{\text{spin}\gamma}(\mathbf{k}) = \sum_{n} f_{n\mathbf{k}} \Omega_{n,\alpha\beta}^{\text{spin}\gamma}(\mathbf{k}).
\end{equation}
The macroscopic SHC tensor ($\sigma_{\alpha\beta}^{\text{spin}\gamma}(\omega)$) is obtained by integrating this $\mathbf{k}$-resolved curvature over the entire Brillouin zone \cite{pizzi2020wannier90,qiao2018calculation}
\begin{equation}
\label{eq:shc}
\sigma_{\alpha\beta}^{\text{spin}\gamma}(\omega) = -\frac{e^2}{\hbar} \frac{1}{V N_k} \sum_{\mathbf{k}} \Omega_{\alpha\beta}^{\text{spin}\gamma}(\mathbf{k}).
\end{equation}
For the analysis of the intrinsic SHC driven by external electrostatic gating, we evaluate the direct current (DC) static limit where the optical frequency $\omega = 0$.

We then investigate the magnetic exchange interactions with the TB2J package \cite{he2021tb2j}, which implements the magnetic force theorem via the Green's function method within the rigid spin-rotation formalism. To explicitly account for the noncollinear magnetic structure and spin-orbit coupling, the energy variation due to two-spin interactions is evaluated by decomposing the Wannier-interpolated Hamiltonian and the corresponding Green's functions into their scalar and vector components ($u, v \in \{0, x, y, z\}$) \cite{he2021tb2j}. The resulting energy variation is captured by the integration matrix $A_{ij}^{uv}$, defined as
\begin{equation}
A_{ij}^{uv} = -\frac{1}{\pi} \int_{-\infty}^{E_F} \text{Tr} \{ p_{i}^{z} G_{ij}^{u} p_{j}^{z} G_{ji}^{v} \} d\epsilon,
\end{equation}
where $G_{ij}$ represents the inter-site Green's function, $p_{i}$ is the intra-atomic Hamiltonian component, and $E_F$ is the Fermi energy. Within this formalism, the isotropic Heisenberg exchange ($J^{\mathrm{iso}}_{ij}$), the anisotropic exchange tensor ($J^{\mathrm{ani}}_{ij}$), and the Dzyaloshinskii-Moriya interaction (DMI) vector ($\mathbf{D}_{ij}$) are systematically extracted from the $A_{ij}$ matrix elements through the following relations \cite{he2021tb2j, szilva2013interatomic}
\begin{align}
J_{ij}^{\mathrm{iso}} &= \text{Im}(A_{ij}^{00} - A_{ij}^{xx} - A_{ij}^{yy} - A_{ij}^{zz}),\\
J_{ij}^{\mathrm{ani},uv} &= \text{Im}(A_{ij}^{uv} + A_{ij}^{vu}),\\
D_{ij}^{u} &= \text{Re}(A_{ij}^{0u} - A_{ij}^{u0}).
\end{align}
These extracted parameters are subsequently mapped onto a generalized spin Hamiltonian
\begin{equation}
\mathcal{H}_{\mathrm{spin}} = - \sum_{i\neq j} \left[ J^{\mathrm{iso}}_{ij} \mathbf{S}_i \cdot \mathbf{S}_j + \mathbf{S}_i^{\mathrm{T}} J^{\mathrm{ani}}_{ij} \mathbf{S}_j + \mathbf{D}_{ij} \cdot (\mathbf{S}_i \times \mathbf{S}_j) \right],
\end{equation}
where $\mathbf{S}_i$ and $\mathbf{S}_j$ represent the unit spin vectors at sites $i$ and $j$, respectively \cite{he2021tb2j}. This Hamiltonian is then employed to compute the macroscopic magnon dispersion, which is detailed in Section~\ref{subsec:magnon}.

\section{Results and Discussion}
\label{sec:results}

\subsection{Electronic Properties}
\label{subsec:electronic}

To understand the underlying electronic behavior of the strain-induced ferromagnetic phase, we calculate the band structures and the corresponding density of states (DOS) along with the projected density of states (PDOS) with spin-orbit coupling (SOC) and noncollinear spin formalism. Figure~\ref{fig:band} presents the electronic structures of the 2.5\% and 5\% strained $\text{CrSiSe}_3$ monolayers under various external electric field perturbations along the $z$-direction. Specifically, the top row [Figs.~\ref{fig:band}(a)-(d)] depicts the evolution of the electronic states for the 2.5\% strained structure under applied electric field amplitudes of 0, 0.1, 0.2, and 0.3 V/\AA, respectively. Similarly, the bottom row [Figs.~\ref{fig:band}(e)-(h)] displays the corresponding states for the 5\% strained structure under the identical field variations.

Our calculations reveal that both the 2.5\% and 5\% strained structures exhibit a stable semiconducting nature, characterized by indirect band gaps, $E_g \approx 0.56$ eV and $E_g \approx 0.64$ eV, respectively. To identify the specific orbital origin governing these electronic states near the band edges, we resolve the total DOS into atomic projected density of states (PDOS). As shown in the rightmost columns of each panel in Figure~\ref{fig:band}, the states near the Fermi level ($E_F = 0$ eV) are predominantly composed of the hybridized Cr-$d$ and Se-$p$ orbitals, with a minor, but still noticeable, contribution from the Si-$p$ orbitals at the valence band top.  On the other hand, other electronic states, including the $s$-orbitals of all constituent atoms and the $d$-orbitals of the selenium atoms, exhibit negligible intensity within the active window around $E_F$ or reside deeply embedded within the core energy levels. Consequently, these minor orbitals do not actively participate in the low-energy electronic interactions, providing a solid justification for selecting only Cr-$d$, Se-$p$, and Si-$p$ orbitals as the localized basis projection during the subsequent Wannierization process.

We find that the $\text{CrSiSe}_3$ monolayers exhibit pronounced electronic stability across different strain configurations. As the external electric field is incrementally applied up to 0.3 V/\AA, neither the magnitudes of the band gaps nor the fundamental dispersions of the valence and conduction bands near the Fermi level undergo significant modulation. For instance, the 0.64 eV band gap of the 5\% strained system remains entirely invariant under all applied field strengths. The total and projected DOS further corroborates this stability, revealing no prominent redistribution, orbital re-hybridization, or shifting of the electronic states near the band edges under the applied fields.  Such a high degree of electronic robustness against external electric fields highlights the potential of the strain-engineered $\text{CrSiSe}_3$ monolayer for practical device integration. In realistic operational environments, quantum materials are often subjected to stray electric fields or gate voltages. The ability of $\text{CrSiSe}_3$ to maintain its intrinsic semiconducting gap, stable electronic states, and invariant orbital composition under such perturbations makes it a reliable candidate for next-generation spintronic applications.

\subsection{Robustness of Charge Topology and Optical Responses}
\label{subsec:topo_optical}

\begin{figure}[t]
    \centering
    \includegraphics[width=\linewidth]{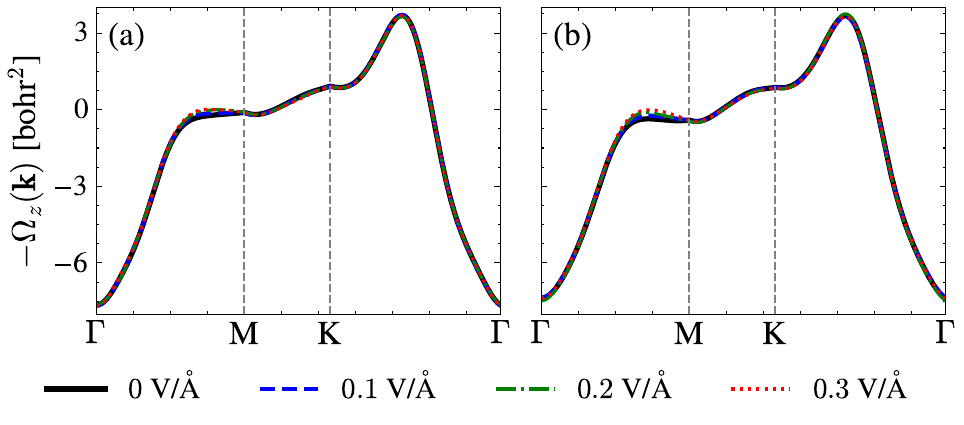}
    \caption{One-dimensional distribution profiles of the charge Berry curvature ($-\Omega_z$) along the high-symmetry $k$-path for the strained $\text{CrSiSe}_3$ monolayers under varying external out-of-plane electric fields. The left panel (a) and right panel (b) illustrate the 2.5\% and 5.0\% strained configurations, respectively. The global legend at the bottom defines the curves for each applied field strength.}
\label{fig:charge_berry}
\end{figure}

Following up the confirmation of the robust electronic band structure in the strained $\text{CrSiSe}_3$ monolayers, we are triggered to investigate whether this structural and electronic resilience extends to its topological features. In 2D ferromagnetic materials, the breaking of time-reversal symmetry, combined with the intrinsic spin-orbit coupling (SOC), often gives rise to non-trivial topological phases. These phases are characterized by a non-zero distribution of the Berry curvature in the momentum space, which acts as an effective magnetic field for the Bloch electrons and governs the anomalous transport behaviors of the system. 

To evaluate these topological characteristics, we compute the intrinsic charge Berry curvature along the high-symmetry $\mathbf{k}$-path via the Wannier interpolation scheme based on the antisymmetric tensor relation detailed in Eq.~\eqref{eq:curv_tensor}. The resulting momentum-space distribution profile serves as a direct fingerprint of the topological nature of the electronic ground state.  As plotted in Fig.~\ref{fig:charge_berry}, the distribution profiles of the charge Berry curvature reveal a highly consistent qualitative behavior across all applied electric field variations. The prominent features of the Berry curvature remain largely stationary in the momentum space, although a closer inspection may display a subtle modulation in the peak magnitudes located along the $\Gamma$--$\text{M}$ path as the electric field increases.  Despite this minor localized variation, the absence of any peak sign reversal or massive momentum shifting provides strong evidence that the fundamental topological invariants of the strained $\text{CrSiSe}_3$ monolayer are intrinsically bound to its robust electronic band structure. Consequently, the out-of-plane electrical perturbations up to 0.3~V/\AA\ are insufficient to induce any topological phase transition or significant band inversion that would otherwise alter the overall $\Omega_z(\mathbf{k})$ dispersion.

\begin{figure}[t]
    \centering
    \includegraphics[width=\linewidth]{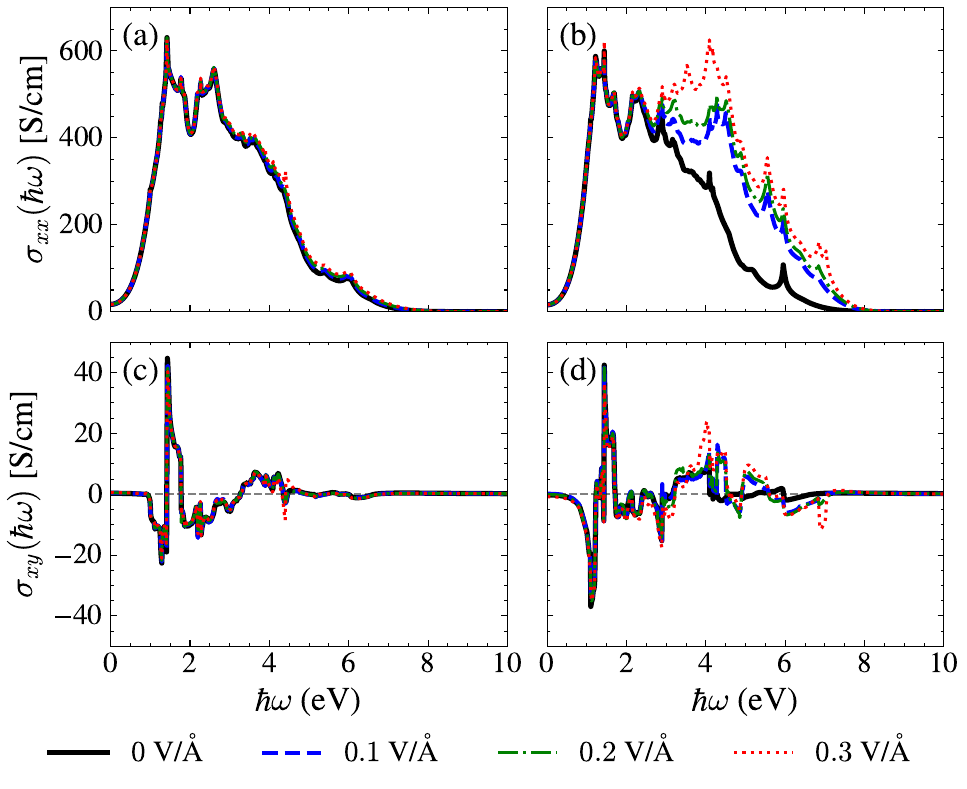}
    \caption{Frequency-dependent optical conductivity spectra of the strained $\text{CrSiSe}_3$ monolayers under various external out-of-plane electric fields. The top row illustrates the symmetric (longitudinal) component, $\sigma_{xx}(\omega)$, for the (a) 2.5\% and (b) 5.0\% strained configurations, while the bottom row displays the antisymmetric (Hall) component, $\sigma_{xy}(\omega)$, for the (c) 2.5\% and (d) 5.0\% strained configurations. The global legend at the bottom defines the curves for each applied field strength.}
    \label{fig:kubo}
\end{figure}

The pronounced stability of the electronic structure and charge topology is further corroborated by macroscopic optical responses. The frequency-dependent optical conductivity tensor, $\sigma_{\alpha\beta}(\omega)$, is evaluated using the Kubo-Greenwood formalism presented in Eq.~\eqref{eq:kubo_general}. By systematically separating the tensor into its symmetric and antisymmetric parts according to Eq.~\eqref{eq:sym_antisym}, we resolve the independent macroscopically observed components, namely the symmetric longitudinal conductivity, $\sigma_{xx}(\omega)$, and the antisymmetric optical Hall conductivity, $\sigma_{xy}(\omega)$. The symmetric component fundamentally governs the standard optical absorption transitions, whereas the antisymmetric part is linked to the non-zero integration of the ground-state topological curvature under finite optical frequencies.

Figure~\ref{fig:kubo} presents the calculated optical conductivity spectra for both 2.5\% and 5.0\% strain configurations under varying out-of-plane electric fields. The top row [Figs.~\ref{fig:kubo}(a) and (b)] displays the symmetric (longitudinal) component, $\sigma_{xx}(\omega)$, which is primarily governed by direct optical transitions between the occupied valence bands and empty conduction bands. The spectra across all electric field variations exhibit a highly consistent profile, reaching a prominent maximum conductivity of approximately 600~S/cm. While the low-energy optical responses are nearly identical, minor deviations emerge under specific conditions: the 5\% strained system exhibits slight spectral variations at higher photon energies ($> 2$~eV), and the 2.5\% strained system shows negligible modifications exclusively under the maximum applied field of 0.3~V/\AA. Nevertheless, the invariance of the primary macroscopic peaks confirms that the fundamental energy spacings and the dipole transition matrix elements are not strongly perturbed by external gating.

Similarly, the bottom row [Figs.~\ref{fig:kubo}(c) and (d)] illustrates the antisymmetric (Hall) component of the optical conductivity, $\sigma_{xy}(\omega)$, which is intimately related to the anomalous Hall effect at optical frequencies and the integration of the Berry curvature over all occupied states. The $\sigma_{xy}$ spectra are virtually indistinguishable across the 0 to 0.3~V/\AA\ field range, consistently exhibiting a distinct peak with a maximum magnitude of around 40~S/cm. The resilience of these charge-based optical and topological signatures decisively establishes that the charge transport mechanisms in the strained $\text{CrSiSe}_3$ monolayers are highly protected against external voltage fluctuations.

\subsection{Tunable Spin Transport and Spin Hall Effect}
\label{subsec:spin}

\begin{figure}[t]
    \centering
    \includegraphics[width=\linewidth]{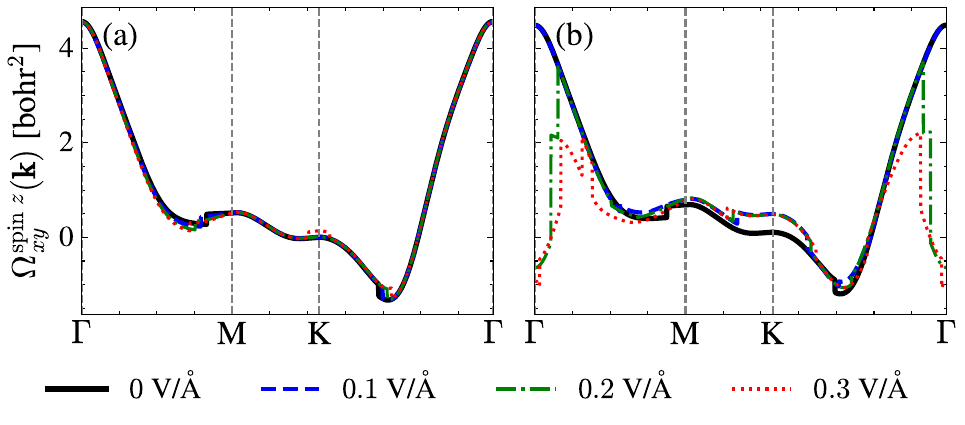}
    \caption{One-dimensional distribution profiles of the spin Berry-like curvature ($\Omega_{xy}^{\text{spin}~z}$) along the high-symmetry $\mathbf{k}$-path for the strained $\text{CrSiSe}_3$ monolayers under varying external out-of-plane electric fields. The left panel (a) and right panel (b) illustrate the 2.5\% and 5.0\% strained configurations, respectively. The global legend at the bottom defines the curves for each applied field strength.}
    \label{fig:spin_berry}
\end{figure}

Modern spintronic applications also demand materials that possess stable charge transport and offer controllable spin degrees of freedom. While the charge-transport features of the strained $\text{CrSiSe}_3$ monolayers are robust under external fields, a critical question remains regarding the responsiveness of their spin-dependent properties. To accurately capture this spin dynamics, we evaluate the intrinsic SHC within the independent-particle approximation using the Kubo-Greenwood formalism. As detailed in the computational methods, the macroscopic spin transport is fundamentally governed by the band-projected spin Berry-like curvature defined in Eq.~\eqref{eq:spin_berry_band}. By integrating the $\mathbf{k}$-resolved spin curvature [Eq.~\eqref{eq:spin_berry_k}] over the Brillouin zone in the DC static limit ($\omega = 0$), the macroscopic SHC tensor [Eq.~\eqref{eq:shc}] is obtained to analyze the responses under varying electrostatic gating.

The resulting distribution of $\Omega_{xy}^{\text{spin}~z}(\mathbf{k})$ along the high-symmetry paths elucidates the topological origins of the spin transport. As shown in Fig.~\ref{fig:spin_berry}, the underlying spin topology exhibits distinct field-induced modulations that depend strongly on the strain configuration. For the 2.5\% strained system, minor localized variations emerge across three specific momentum regions, namely along the $\Gamma$--M path, at the high-symmetry K point, and along the K--$\Gamma$ segment. These subtle features are initiated even at a relatively low field strength of 0.1~V/\AA. In stark contrast, the 5\% strained configuration displays highly significant modifications under external gating. Most notably, under elevated field strengths of 0.2 and 0.3~V/\AA, the spin Berry-like curvature at the $\Gamma$ point undergoes a field-induced sign reversal, flipping from its initial positive values into a distinctly negative regime. This field-driven sign reversal in the microscopic spin topology, alongside the localized peak modulations, serves as the fundamental mechanism underpinning the sensitive, non-monotonic tunability observed in the macroscopic SHC.

\begin{figure}[t]
    \centering
    \includegraphics[width=\linewidth]{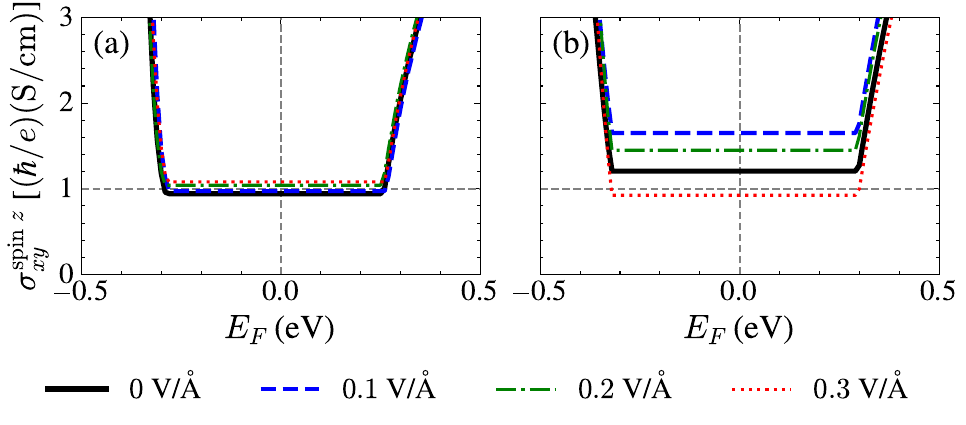}
    \caption{Intrinsic spin Hall conductivity (SHC) as a function of Fermi energy for the strained $\text{CrSiSe}_3$ monolayers under various out-of-plane external electric fields. The left panel (a) and right panel (b) illustrate the SHC spectra for the 2.5\% and 5.0\% strained configurations, respectively. The global legend defines the curves for each applied field strength. The horizontal gray dashed line in each panel denotes a reference SHC value of 1~$(\hbar/e)(\text{S}/\text{cm})$ to clearly visualize the field-induced transport enhancement.}
    \label{fig:shc}
\end{figure}

The energy-integrated results for the macroscopic transport, presented in Fig.~\ref{fig:shc}, confirm the topological sensitivity and they also reveal a fascinating tunability of the SHC under applied electric fields. For the 2.5\% strained configuration [Fig.~\ref{fig:shc}(a)], the SHC exhibits a clear and steady amplification trend. In the absence of an electric field, the stable plateau of the SHC lies entirely below the reference line of 1~$(\hbar/e)(\text{S}/\text{cm})$. As the field increases, the plateau successfully breaches and sustains values well above this reference magnitude.  On the other hand, the 5\% strained system [Fig.~\ref{fig:shc}(b)] demonstrates a distinct, non-monotonic spin response. The application of a 0.1~V/\AA\ field induces a dramatic amplification of the SHC plateau compared to its pristine state. However, further increasing the field to 0.2~V/\AA\ slightly attenuates the magnitude, though it remains elevated relative to the baseline. Ultimately, a strong field of 0.3~V/\AA\ severely suppresses the SHC, driving it to its lowest value across all variations.

Despite these dynamic modulations in peak magnitudes, the energy width of the SHC plateaus remains remarkably constant across all electric field variations for both strain configurations.  Such a persistent energy window is consistent with the previously established electronic robustness of the system.  These phenomena highlight contrasting response in the strain-engineered $\text{CrSiSe}_3$ monolayer, where its fundamental electronic states, band gap, and charge topology are robust against external perturbations, yet its spin transport characteristics can be effectively tuned and modulated via external electric field gating. 

\subsection{Magneto-Optical Properties}
\label{subsec:moke}

\begin{figure}[t]
    \centering
    \includegraphics[width=\linewidth]{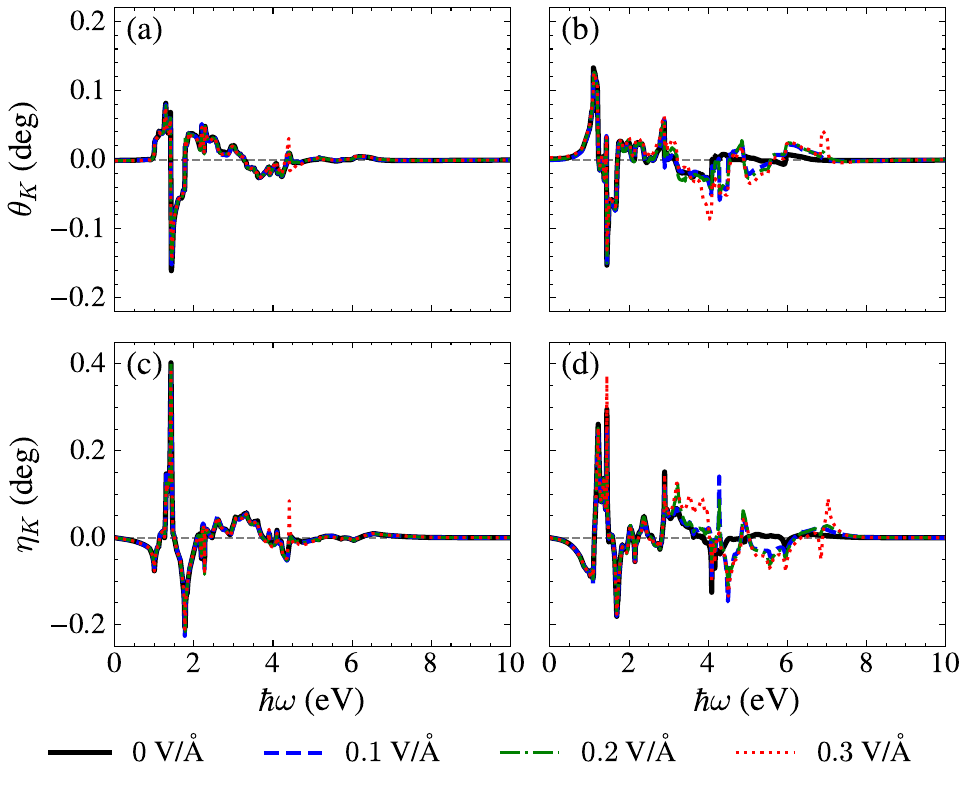}
    \caption{Frequency-dependent magneto-optical Kerr rotation ($\theta_K$) and ellipticity ($\eta_K$) for the strained $\text{CrSiSe}_3$ monolayers under various out-of-plane external electric fields. The top row [panels (a) and (b)] illustrates the $\theta_K$ spectra for the 2.5\% and 5.0\% strained configurations, respectively, under electric fields of 0, 0.1, 0.2, and 0.3~V/\AA. The bottom row [panels (c) and (d)] displays the corresponding $\eta_K$ responses for the identical strained systems under the same field variations. The horizontal gray dashed line in each panel denotes a reference value of 0~deg to clearly visualize the vanishing magneto-optical response in the static DC limit.}
    \label{fig:moke}
\end{figure}

To complete the optical characterization and bridge it with the magnetic properties of the $\text{CrSiSe}_3$ monolayer, we also calculate the macroscopic magneto-optical responses, specifically the Kerr rotation angle ($\theta_K$) and Kerr ellipticity ($\eta_K$). As detailed in the computational methods, the complex Kerr spectra under normal light incidence can be derived directly from the calculated 2D antisymmetric Hall conductivity, $\sigma_{xy}^{\text{2D}}(\hbar\omega)$.  This evaluation assumes the monolayer resides on a standard insulating dielectric substrate with a representative refractive index of $n=2$ \cite{pratama2020circular,adhidewata2023trigonal}.

The calculated frequency-dependent Kerr spectra for both 2.5\% and 5.0\% tensile strains under various out-of-plane electric fields are presented in Fig.~\ref{fig:moke}. For the 2.5\% strained configuration, the Kerr rotation ($\theta_K$) is basically unaltered up to a photon energy of $\hbar\omega \approx 4$~eV, with a distinct variation emerging exclusively under the highest electric field of 0.3~V/\AA.  Meanwhile, for the Kerr ellipticity ($\eta_K$), noticeable deviations among the different applied fields begin to appear above $\hbar\omega \approx 2$~eV. Upon increasing the tensile strain to 5.0\%, the Kerr rotation displays a remarkably similar trend to that of the 2.5\% case. In contrast, the Kerr ellipticity for the 5.0\% strained system shows an earlier onset of field-induced modifications starting around $\hbar\omega \approx 1.5$~eV, particularly for the 0.3~V/\AA\ case, followed by minor deviations among all field variations at higher energies. These detailed spectral features are intimately linked to the subtle modifications of the interband transition channels under the combined influence of strain and electrostatic gating.

Consistent with the longitudinal and Hall optical conductivities, the magneto-optical spectra exhibit a substantial degree of robustness against external electric fields. As depicted in Fig.~\ref{fig:moke}, the primary Kerr resonance peaks remain stable in both magnitude and energy position, particularly in the lower energy regime and for field strengths up to 0.2~V/\AA. Although minor field-induced deviations emerge at higher photon energies and under the maximum applied field of 0.3~V/\AA, the overall resilient macroscopic behavior implies that the $\text{CrSiSe}_3$ monolayer is suitable for stable magneto-optical readout applications. Its optical response provides a reliable signal that is highly resistant to perturbations from typical stray ambient electric fields.

\subsection{Magnonic Properties}
\label{subsec:magnon}

\begin{table}[tb]
\centering
\caption{Average isotropic exchange interactions ($J_{\text{iso}}$) for the $\text{CrSiSe}_3$ monolayer at various $\text{Cr}$--$\text{Cr}$ bond distances under different out-of-plane external electric fields, comparing the 2.5\% and 5.0\% tensile strain configurations.}
\begin{tabular}{l c c c c c}
\toprule
\multirow{2}{*}{\textbf{Neighbor}} & \multirow{2}{*}{\textbf{Distance (\AA)}} & \multicolumn{4}{c}{\textbf{$J_{\text{iso}}$ (meV) at $E$-fields (V/\AA)}} \\
\cmidrule(lr){3-6}
 & & \textbf{0.0} & \textbf{0.1} & \textbf{0.2} & \textbf{0.3} \\
\midrule
\multicolumn{6}{l}{\textit{2.5\% Tensile Strain}} \\
NN   & 3.72 &  3.28 &  3.14 &  3.35 &  3.29 \\
NNN  & 6.44 &  0.35 &  0.28 &  0.36 &  0.34 \\
NNNN & 7.44 & -1.23 & -1.32 & -1.23 & -1.25 \\
\midrule
\multicolumn{6}{l}{\textit{5.0\% Tensile Strain}} \\
NN   & 3.81 &  4.59 &  4.82 &  5.12 &  4.87 \\
NNN  & 6.60 &  0.20 &  0.29 &  0.39 &  0.30 \\
NNNN & 7.62 & -1.21 & -1.14 & -1.10 & -1.15 \\
\bottomrule
\end{tabular}
\label{tab:j_iso}
\end{table}

\begin{figure*}[t]
    \centering
    \includegraphics[width=\linewidth]{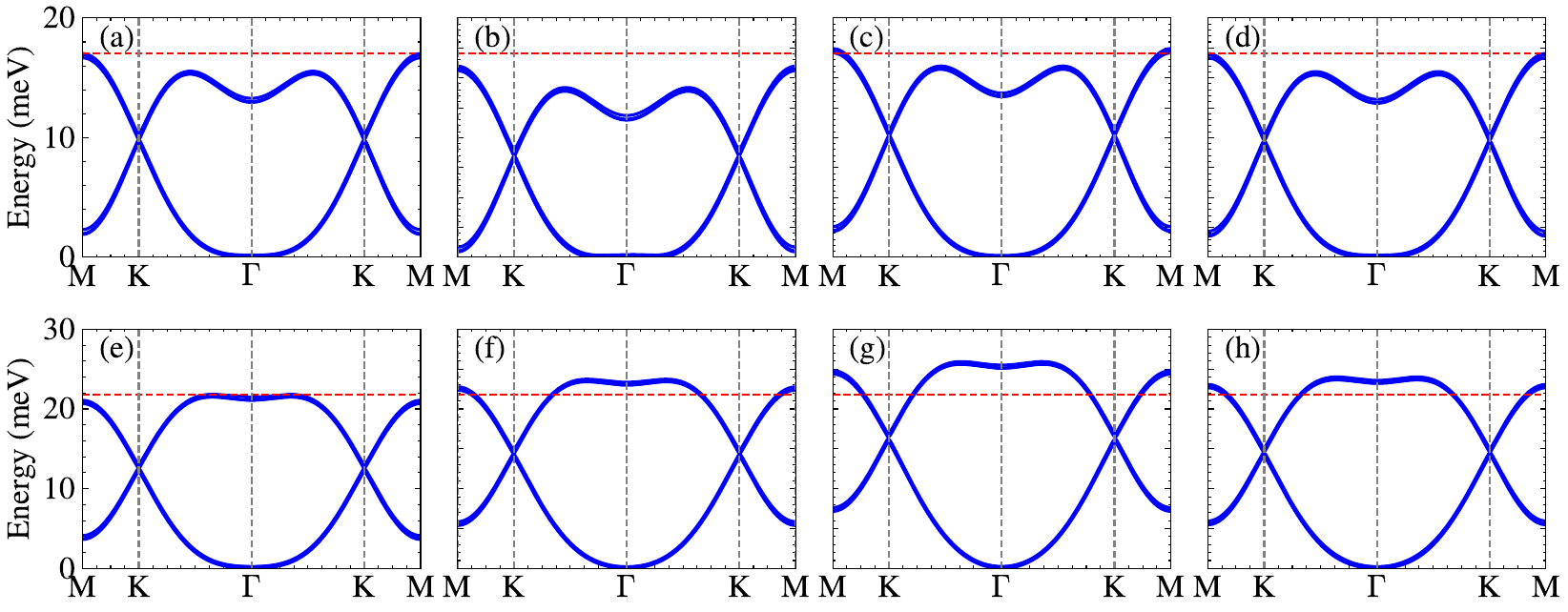}
    \caption{Magnon dispersion spectra of the strained $\text{CrSiSe}_3$ monolayers across the Brillouin zone under various external out-of-plane electric fields. The top row [panels (a)--(d)] illustrates the non-monotonic evolution of the 2.5\% strained configuration under 0, 0.1, 0.2, and 0.3 V/\AA, respectively. The bottom row [panels (e)--(h)] illustrates the corresponding magnon dispersions for the 5.0\% strained system under the identical field variations.}
    \label{fig:magnon}
\end{figure*}

To gain deeper insights into the dynamic spin response and the magnetic stability of the system under external perturbations, we calculate the magnon dispersion, which describes the elementary collective excitations of the spin waves. The computed magnon spectra across the Brillouin zone are presented in Fig.~\ref{fig:magnon}, comparing the 2.5\% and 5.0\% tensile strain configurations. Remarkably, the application of a larger tensile strain of 5.0\% induces a substantial upward shift in the overall spin-wave excitation energies compared to the 2.5\% variation. This macroscopic amplification is directly rooted in the strengthening of the underlying isotropic exchange interactions ($J_{\text{iso}}$), the details of which are systematically documented in Table~\ref{tab:j_iso}.

Unlike the resilient electronic band structure discussed previously, the magnon dispersions for both strain configurations exhibit distinct, non-monotonic sensitivities to the applied out-of-plane electric field, albeit following entirely different evolution pathways. For the 2.5\% strained system [Figs.~\ref{fig:magnon}(a)-(d)], the pristine state serves as our baseline. When a moderate electric field of 0.1 V/\AA\ is applied, the maximum peak of the magnon dispersion experiences a noticeable reduction. Intriguingly, as the field increases to 0.2 V/\AA, the dispersion peak sharply rises and reaches its highest energy profile, successfully surpassing the zero-field maximum. Upon further increasing the field strength to 0.3 V/\AA, the magnon peak is suppressed once again, returning to an energy level that is nearly identical to its initially unperturbed configuration. 

In a stark contrast, the 5.0\% strained configuration [Figs.~\ref{fig:magnon}(e)-(h)] demonstrates a highly unique field-driven evolution. Starting from a higher pristine energy baseline of approximately 21~meV at zero field, the maximum magnon excitation energy undergoes a continuous amplification when electric fields of 0.1 and 0.2~V/\AA\ are sequentially introduced, reaching its absolute peak at 0.2~V/\AA. Upon reaching the maximum field intensity of 0.3~V/\AA, the magnon dispersion peak experiences an attenuation, but it crucially remains elevated and does not drop as low as the energy level observed at 0.1~V/\AA. This behavior confirms that the collective spin excitations in the 5.0\% strained monolayer maintain a tunable profile under electrical gating.

The fundamental mechanism driving these dynamic modulations can be elucidated by examining the calculated $J_{\text{iso}}$ parameters listed in Table~\ref{tab:j_iso}. Within the framework of the Heisenberg model, the magnon dispersion and the associated energy excitation scales directly with the coupling strengths between adjacent magnetic moments, particularly the nearest-neighbor (NN) interactions. For the 5.0\% strain configuration, the NN exchange constant rises from 4.59~meV at zero field to a maximum of 5.12~meV at 0.2~V/\AA, before decreasing to 4.87~meV at 0.3~V/\AA. This non-monotonic evolution of the exchange coupling dictates the field-sensitive shift in the magnon excitation energies. The fluctuations suggest that the external out-of-plane electric field actively modulates the Cr--Se--Cr superexchange pathways, likely by inducing a spatial polarization of the intermediate Se electron clouds that mediate the magnetic interactions between neighboring Cr atoms.

The calculated isotropic exchange parameters ($J_{\text{iso}}$) are summarized in Table~\ref{tab:j_iso}. We observe that the ferromagnetic ground state is primarily governed by the nearest-neighbor isotropic exchange interaction, $J_{\text{iso}} \approx 3.10$ meV. The relativistic DMI ($D$) and anisotropic exchange ($J_{\text{ani}}$) terms are also computed. These magnitudes remain smaller than $J_{\text{iso}}$ by roughly two orders of magnitude and the full set of these parameters are provided in a public repository~\cite{github}. Nevertheless, these relativistic terms are essential for defining the magnetic anisotropy and stabilizing the spin orientation in the $\text{CrSiSe}_3$ monolayer.

Up to this stage, we find that the magnonic properties complement our findings regarding the topological spin responses. A profound physical picture emerges for the strained $\text{CrSiSe}_3$ monolayers, as their charge-driven electronic band gaps, total density of states, and fundamental charge topology remain immune to out-of-plane electrical perturbations, whereas their spin transport signatures and microscopic magnon excitations demonstrate a controllable tunability. This distinctive decoupling between charge robustness and spin responsiveness establishes the strain-engineered $\text{CrSiSe}_3$ monolayer as a promising candidate for electric-field-controlled spintronic and magnonic field-effect devices.

\section{Conclusions}
\label{sec:conclusion}

We have investigated the electronic, topological, magnonic, and magneto-optical properties of strain-engineered ferromagnetic $\text{CrSiSe}_3$ monolayers under out-of-plane external electric fields using first-principles calculations. Our results uncover a unique distinction of  electrostatic robustness in the electronic charge characteristics alongside a highly responsive and controllable tunability in the spin degrees of freedom.  While the fundamental band gaps, density of states, intrinsic charge Berry curvatures, and frequency-dependent magneto-optical Kerr effect spectra are resilient against electrical perturbations up to 0.3~V/\AA, the spin-dependent properties exhibit significant field-driven modulations. Specifically, the macroscopic spin Hall conductivity displays field-induced transport enhancement, and the underlying magnon dispersion profiles can be effectively tuned via electrostatic gating through the modulation of the microscopic isotropic Heisenberg exchange interactions ($J_{\text{iso}}$).  The unique coexistence of robust charge stability and flexible spin manipulation suggests that the strained $\text{CrSiSe}_3$ monolayer could emerge as a promising candidate for next-generation, energy-efficient spintronic, magnonic, and magneto-optical field-effect devices.

\section*{Data and code availability}
The raw/processed data and codes required to reproduce these findings are available to download from \url{https://github.com/solihinn17/topo-magnonics}.

\section*{CRediT authorship contribution statement}
\textbf{S.Solihin:} Conceptualization, Data Curation, Formal analysis, Investigation, Methodology, Software, Visualization,  Writing--original draft. \textbf{A.R.T.Nugraha:} Conceptualization, Formal analysis, Funding acquisition, Investigation, Methodology, Supervision, Validation, Writing--review \& editing. \textbf{M.A.Majidi:} Funding acquisition, Project administration, Supervision, Validation, Writing--review \& editing.

\section*{Declaration of Interests}
The authors declare that they have no known competing financial interests or personal relationships that could have appeared to influence the work reported in this paper.

\section*{Acknowledgments}
We thank QuasiLab and Mahameru BRIN for their minicluster and HPC facilities.  S.S. is supported by a research assistantship and Degree-by-Research scholarship from the BRIN Directorate for Talent Management.

\biboptions{sort&compress}
\bibliography{references}
\end{document}